\documentclass{rspublic}
\usepackage{graphicx}
\usepackage{verbatim}

\begin{document}

\title[Entropic transport]{
Entropic transport - A test bed for the Fick-Jacobs approximation}
\author[P.S. Burada]{P. Sekhar Burada, Gerhard Schmid and Peter H\"anggi}
\affiliation{Institut f\"ur Physik,
  Universit\"at Augsburg,
  Universit\"atsstr. 1,
  D-86135 Augsburg, Germany}

\label{firstpage}

\maketitle

\begin{abstract}{Brownian motion; Entropic transport; Fick-Jacobs equation}
  Biased diffusive transport of Brownian particles through
  irregularly shaped, narrow confining quasi-one-dimensional
  structures is investigated. The complexity of the higher dimensional
  diffusive dynamics is reduced by means of the so-called Fick-Jacobs
  approximation, yielding an effective one-dimensional stochastic
  dynamics. Accordingly,  the elimination of transverse, equilibrated
  degrees of freedom stemming from geometrical confinements and/or
  bottlenecks cause entropic potential barriers which the particles
  have to overcome when moving forward noisily. The applicability and
  the validity of the reduced kinetic description is tested by
  comparing the approximation with Brownian dynamics simulations in
  full configuration space. This non-equilibrium transport in such
  quasi-one-dimensional irregular structures implies for
  moderate-to-strong bias a characteristic violation of the
  Sutherland-Einstein fluctuation-dissipation relation.
\end{abstract}

\section{Introduction}

Diffusion of Brownian particles through narrow, tortuous
confining structures such as micro- and nano-pores, zeolites,
biological cells and microfluidic devices plays a prominent  role in
the dynamical characterization of these systems (Barrer, 1978;
Berezhkovskii \& Bezrukov, 2005; Hille, 2001; 
Kettner {\it et al.}, 2000; Liu {\it et al.}, 1999; 
Matthias \& M\"uller, 2003; M\"uller {\it et al.}, 2000; 
Nixon \& Slater, 2002; Siwy {\it et al.}, 2005; Volkmuth \& Austin, 1992).
Effective control schemes for transport in
these systems requires a detailed understanding of the diffusive
mechanisms involving small objects and, in this regard, an operative
measure to gauge the role of fluctuations. The study of these
transport phenomena is in many respects equivalent to an
investigation of geometrically constrained Brownian dynamics 
(Burada {\it et al.}, 2009; H\"anggi \& Marchesoni, 2009; Mazo, 2002). 
With this work we focus on the stochastic transport of small sized
particles in confined geometries and the feasibility of the
so-called {\it Fick-Jacobs} (FJ) approximation to describe the
steady-state particle densities. Restricting the volume of the
configuration space available to the diffusing particles by means of
confining boundaries or obstacles discloses intriguing entropic
phenomena (Liu {\it et al.}, 1999).

The driven transport of charged particles across bottlenecks (Burada
{\it et al.}, 2009), such as ion transport through artificial
nanopores or artificial ion pumps 
(Kosinska {\it et al.}, 2008; Siwy {\it et al.}, 2005) 
or in biological channels (Berezhkovskii \& Bezrukov, 2005) 
are more familiar systems  where diffusive transport
is regulated by entropic barriers. Similarly, the operation of
artificial Brownian motors and molecular machines  relies as well on
a mutual interplay amongst  diffusion and binding action by
energetic or, more relevant in the present context, entropic
barriers (Astumian \& H\"anggi, 2002;
Burada {\it et al.}, 2009; Derenyi \& Astumian, 1998;
H\"anggi \& Marchesoni, 2009; Reimann \& H\"anggi, 2002).

The outline of this work is as follows: In Sec.~2 we introduce our
model and formulate the mathematical formalism needed to model the
diffusion of a Brownian particle immersed in a confined medium. In
Sec.~3 we present the FJ approximation and compute  the entropic
effects on the particle transport and on the steady-state
probability density in the presence of an applied force in
transport direction. In Sec.~4 we compare the numerical precise 2D
simulation results with those obtained from applying the
FJ approximation.  In Sec.~5 we discuss the effective
lateral diffusion and test the Sutherland-Einstein fluctuation-dissipation
relation. Sec.~6 provides a discussion of our main findings.

\section{Overdamped system dynamics}

\begin{figure}[ht]
  \centering
  \includegraphics{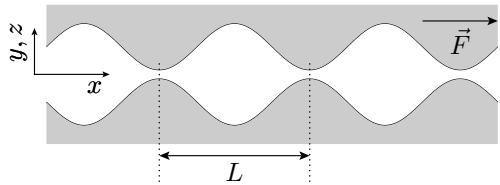}
  \caption{Schematic diagram of a channel confining the
   motion of laterally forced (with strength $F$)Brownian particles.
   The half-width $\omega$ is a periodic function of
   $x$ with periodicity $L$.}
\label{fig:tube}
\end{figure}

Generic mass transport  through confined structures such
as irregular pores and channels, c.f.  the one depicted with
Fig.~\ref{fig:tube}, is governed by the transport of suspended
Brownian particles subjected to an externally applied potential
$V(\vec r)$. Generally, the dynamics of the Brownian particle inside
the medium can  be well described by a Langevin dynamics in the
over-damped limit (Purcell, 1977), with reflecting boundary
conditions at the channel walls. The stochastic dynamics then reads

\begin{align}
\label{eq:Ent-langevin}
\eta \dot{\vec r}(\tilde{t}) =
-\vec{\nabla}V(\vec r(\tilde{t}))+
\sqrt{\eta \, k_{\rm B}\,T}\, \vec{\xi}(\tilde{t})\, ,
\end{align}
where $\vec r$ is the position vector of a Brownian particle at time
$\tilde{t}$, $\eta$ denotes the friction coefficient, $k_{\rm B}$ is
the Boltzmann constant and $T$ refers to the environmental
temperature. Thermal fluctuations due to the coupling of the
Brownian particle to the environment are modeled by Gaussian white
noise with zero mean and an auto-correlation function obeying the
Sutherland-Einstein fluctuation-dissipation relation (H\"anggi \& Marchesoni, 2005):
\begin{align}
\label{eq:G-delta}
\langle \xi_{i}(\tilde{t})\,\xi_{j}(\tilde{t}') \rangle =
2\, \delta_{ij}\, \delta(\tilde{t} - \tilde{t}') \,\,\,
\text{for}\,\, i,j = x,y,z \, .
\end{align}

For simplicity we consider the dynamics of a Brownian particle that
is subjected to an constant force $\vec{F} = F \vec{e}_{x}$ acting
along the direction of the channel axis (here in $x$-direction). The
Langevin equation for the over-damped dynamics then reads:
\begin{align}
  \label{eq:Ent-langevin1}
  \eta\, \dot{\vec r}(\tilde{t}) = \vec{F} +
  \sqrt{\eta \, k_\mathrm{B}\,T}\, \vec{\xi}(\tilde{t}) \, ,
\end{align}
with reflecting (i.e. no across flow) boundary conditions implied at
the channels walls which confine the Brownian particles within the
channel geometry.

In order to further simplify the treatment of this set up we
introduce dimensionless variables. We  measure all lengths in units
of the period length $L$, i.e., $\vec{r} = L \vec{x}$, where
$\vec{x}$ denotes the dimensionless position vector of the particle.
As the unit of time $\tau$ we choose twice the time the particle
takes to diffusively overcome the distance $L$, which is given by
$\tau = L^{2}\eta /(k_{\rm B}T)$, i.e. $\tilde{t} = \tau t$ (Burada
{\it et al.}, 2008$b$). In these dimensionless variables the Langevin
dynamics assumes the form
\begin{align}
  \label{eq:langevin}
  \frac{\mathrm{d}\vec{x}}{\mathrm{d}t } = \vec{f} +
  \vec{\xi}(t)\, .
\end{align}
where $\langle \xi (t) \rangle = 0$,
$\langle \xi_{i}(t)\,\xi_{j}(t') \rangle = 2\, \delta_{ij}\,
\delta(t - t')$ for $i,j = x,y,z$, and the dimensionless force becomes
\begin{equation}
\vec{f} = f \vec{e}_{x} \quad \text{and} \; f= \frac{LF}{k_{\rm B}T} \, .
\label{f}
\end{equation}
The performed dimensionless scaling parameter $f$ characterizes the
force as {\it the ratio between the work  $LF$ done on the particle
along a distance of the period length $L$ and the thermal energy
$k_{\mathrm{B}}T$.} We anticipate here the fact that in the case of diffusion
occurring in purely energetic potential landscapes the driving force
$F$ and the temperature $T$ are independent variables; in contrast,
in systems with entropic features these two quantities become
coupled (Reguera {\it et al.} 2006). In order to adjust a certain
value of $f$ one can modify either the force strength $F$ or adjust
the noise intensity $k_{\mathrm{B}}T$.

The corresponding Fokker-Plank equation describing the time
evolution of the probability density $P(\vec x, t)$ takes the
form (H\"anggi \& Thomas, 1982; Risken, 1989):
\begin{subequations}
\begin{align}
\label{eq:fp}
\frac{\partial P(\vec x, t)}{\partial t} = - \vec{\nabla} \cdot
\vec{J}(\vec{x},t)\, ,
\end{align}
where $\vec J(\vec x, t)$ is the probability current:
\begin{align}
  \label{eq:particlecurrent}
  \vec J(\vec x, t) =
  \left( \vec{f} - \vec{\nabla}\right) P(\vec x, t) \, .
\end{align}
\end{subequations}
\noindent Note that, for channels with similar geometry which are
related by a scale transformation $\vec{r} \to \lambda \vec{r}$,
$\lambda >0$, the transport properties are determined by a single
dimensionless parameter $f$ which subsumes the respective period
length, the external force and the temperature of the surrounding
fluid.

The no-flow condition beyond  the channel walls leads to a vanishing
probability current at those boundaries. Consequently, due to the
impenetrability of the channel walls, the normal component of the
probability current $\vec J(\vec x, t)$ vanishes at those
boundaries. Thus, the boundary conditions at the channel walls are
given by

\begin{align}
  \label{eq:bc-general}
  \vec J(\vec x, t) \cdot \vec{n}=0\; , \quad \vec{x} \in \text{channel
  wall}\,,
\end{align}
where $\vec{n}$ denotes the normal vector at the channel walls. \\

The boundary of a 2D periodic channel, which is mirror symmetric
about the  $x$-axis, is given by the dimensionless periodic
functions $y= \pm \omega(x)$, i.e., $\omega(x+1)=\omega(x)$ for all
$x$, where $x$ and $y$ are the
 cartesian components of $\vec{x}$. In this case, the boundary
condition reads

\begin{align}
  \label{eq:bc}
  \frac{\mathrm{d} \omega (x)}{\mathrm{d} x}
 \left[ f P(x,y,t) - \frac{\partial P(x,y,t)}{\partial
     x} \right] 
  + \frac{\partial P(x,y,t)}{\partial y} = 0\, ,
  \end{align}
at $y=\pm \omega (x)$. Except for a straight channel with $\omega =
\text{\it const}$, there are no periodic channel shapes for which an
exact analytical solution of the Fokker-Planck equation
(\ref{eq:fp}, \ref{eq:particlecurrent}) with the elaborate boundary
conditions in (\ref{eq:bc}) is presently known. Approximate
solutions though can be obtained on the basis of a one-dimensional
diffusion problem proceeding in an effective potential. Narrow
channel openings, which act as geometric hindrances in the original
system, then manifest themselves as entropic barriers within an
effective one-dimensional diffusive FJ approximation 
(Burada {\it et al.}, 2007; Jacobs, 1967; Kalinay \& Percus, 2006;
Reguera {\it et al.}, 2006; Reguera \& Rubi, 2001; Zwanzig, 1992).

\section{Equilibration in transverse channel directions:
the Fick-Jacobs approximation}

In the absence of an external force, i.e. for $\vec{f}=
0$, it was shown (Jacobs, 1967; Kalinay \& Percus, 2006;
Reguera \& Rubi, 2001; Zwanzig, 1992) that the dynamics of Brownian
particles in confined structures (such as the one depicted in
Fig.\ref{fig:tube}) can be described approximatively by the
FJ equation; i.e.,

\begin{align}
\label{eq:fickjacobs}
\frac{\partial}{\partial t} P(x,t) =
\frac{\partial}{\partial x} \,D(x) \,e^{-A(x)}
\frac{\partial}{\partial x} e^{A(x)} P(x,t) \, .
\end{align}
This 1D equation is obtained from the full 2D Smoluchowski equation
upon the elimination of the transverse $y$  spatial coordinate
degree of freedom by assuming a much faster equilibration in that
channel direction than in the longitudinal one. 
An analogous reduction mechanism has been used for 
the transport of neutrons through nuclear reactors 
(Beckurts \& Wirtz, 1964).
In the equation (\ref{eq:fickjacobs}), 
$P(x,t) = \int_{-\omega(x)}^{\omega(x)} P(x,y,t)\,{\rm d}y$ denotes 
the marginal probability density along the axis of the channel. $A(x)$
corresponds to the potential of mean force which equals for the
considered situation the free energy, i.e. $A(x)=E(x)-S(x)= 0 - \ln
\omega(x)$. We note that for three dimensional channels an analogue
approximate Fokker-Planck equation holds in which the function
$\omega(x)$ is to be replaced by $\pi \omega^{2}(x)$ (i.e. the area
of the corresponding cross-section). In the original work by Jacobs
(Jacobs, 1967) the 1D diffusion coefficient $D(x)$ is constant and
equals the bare diffusion constant which assumes unity in  present
dimensionless variables. However, introducing the $x$-dependent
diffusion coefficient considerably improves the accuracy of the
kinetic equation, extending its validity to more winding structures
(Burada {\it et al.}, 2007; Reguera \& Rubi 2001; Zwanzig, 1992).
The expression for $D(x)$ reads (in dimensionless units)
\begin{align}
  \label{eq:diffusionconst}
  D(x)=\frac{1}{(1+\omega'(x)^{2})^{\alpha}}\, ,
\end{align}
where $\alpha=1/3,1/2$ for two and three dimensions, respectively,
has been  shown to appropriately account for the curvature effects
of the confining walls (Burada {\itshape et al.}, 2008$b$; 
Reguera \& Rubi, 2001). 
$\omega'(x)$ indicates the first derivative of the
boundary function $\omega(x)$ with respect to $x$.

\begin{figure}[t]
\centering
 \includegraphics{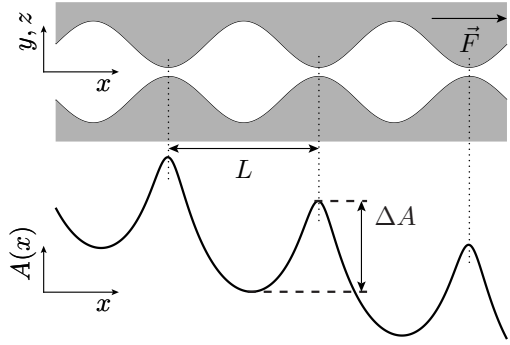}
  \caption{Sketch of the 2D channel and the effective one-dimensional
    potential: the Fick-Jacobs (FJ) approximation allows for a reduction of
    the 2D Brownian dynamics within the periodic channel (periodicity:
    $L$) to an approximate 1D Brownian dynamics with an effective potential which
    is given by the free energy function $A(x)$. In the presence of
    an applied bias $A(x)$ has the form of a tiled periodic potential with
    a barrier height of $\Delta A$ which depends on the temperature $T$.}
\label{fig:free-energy}
\end{figure}

In the presence of a constant force $F$ along the direction of the
channel the FJ diffusion equation (\ref{eq:fickjacobs}) can be
recast into the form (Burada {\it et al.}, 2007; 
Burada {\it et al.}, 2008$b$, Reguera {\it et al.}, 2006):

\begin{align}
\label{eq:fickjacobs_ours}
\frac{\partial P}{\partial t}=\frac{\partial}{\partial
  x}D(x)\left(\frac{\partial P}{\partial
    x}+ \frac{\mathrm{d}A(x)}
    {\mathrm{d}x}P\right)\,
\end{align}
with the dimensionless free energy $A(x) := E(x) - S(x) = -f\, x -
\ln \omega(x)$. In  terms of the original unscaled physical
variables the energy is $\tilde{E} \equiv k_{\rm B}T E(x) = -F
\tilde{x}$ ($\tilde{x} = x L$) and the dimensional entropic
contribution reads $\tilde{S} \equiv k_{\rm B} T S(x) =
k_\mathrm{B} T \,\ln \omega(x)$. For a periodic channel
arrangement this free energy assumes the form of a tilted periodic
potential, see Fig.~\ref{fig:free-energy}. In the absence of a force
the free energy is purely entropic and
Eq.~(\ref{eq:fickjacobs_ours}) reduces to the FJ
equation~(\ref{eq:fickjacobs}). On the other hand, for a straight
channel the entropic contribution vanishes and the particles are
solely driven by the externally applied  force.

Remarkably, the temperature $T$ dictates the strength of the
effective potential. An increase in temperature causes an increase
in barrier height $\Delta A$ while for purely energetic systems the
barrier height is independent of the temperature (H\"anggi {\it et
al.}, 1990).

\subsection{Steady-state probability density}

Formally, the steady-state density of the particles
is obtained in the limit $t \to \infty$, i.e. $P^{\mathrm{st}}(x) =
\lim_{t \to \infty} P(x,t)$. As a consequence,
$\frac{\partial}{\partial t} P^{\mathrm{st}}(x) = 0$. An expression
for the steady-state density can be derived from
Eq.~(\ref{eq:fickjacobs}), using arguments  detailed in  the Appendix.
Using the main result in Eq.~(\ref{eq:steadystate}) one obtains
\begin{align}
  \label{eq:probdist}
  P^{\mathrm{st}}(x) & = \frac{I(x,f)}{\displaystyle \int_{0}^{1} I(z,f)
    \, \mathrm{d}z}\, ,
  \intertext{where}
  \label{eq:integrals}
  I(x, f) & = e^{-A(x)}\,\mathrm{d}x
    \displaystyle \int_{x}^{x+1}\,\frac{e^{A(x')}}{D(x')}
    \,\mathrm{d}x' \, ,
\end{align}
depends on the dimensionless position $x$, the force $f$ and via the
position dependent diffusion coefficient on the shape of the tube
given in terms of the shape function $\omega(x)$ and its first
derivative, cf. Eq.~(\ref{eq:diffusionconst}). Note, that the
probability density $P^{\mathrm{st}}(x)$ is normalized on the
unit interval.

\subsection{Nonlinear mobility}

The primary quantity of particle transport through
periodic channels is the average {\it particle current}, $\langle
\dot{x} \rangle$, or equivalently the {\it nonlinear mobility},
which is defined as the ratio between  the average particle current
and the applied force $f$. For the average particle current we
derive an expression which is similar to the Stratonovich formula
for the current occurring in titled periodic energy landscapes, but here with a spatially
dependent diffusion coefficient (Burada {\it et al.}, 2007). A detailed derivation of this
expression is given in the Appendix~(\ref{app:steadystate}), cf.
Eq.~(\ref{eq:j8}). Hence, we obtain the nonlinear mobility for a 2D or 3D
channel:

\begin{align}
  \label{eq:nonlinearmobility}
  \mu(f) = \frac{\langle \dot{x} \rangle}{f} =
\frac{1}{f}\, \frac{1-e^{-f}}{{\displaystyle \int_{0}^{1}}\,
      I(z, f)\,\mathrm{d}z} \, ,
\end{align}
with $I(z, f)$ given in Eq.~(\ref{eq:integrals}).

\section{Precise numerics for a 2D channel geometry}

The steady-state density and the average particle
current, predicted analytically above, has been  compared with
Brownian dynamic simulations performed by a numerical integration of
the Langevin equation Eq.~(\ref{eq:langevin}), using the stochastic
Euler-algorithm. The shape of the exemplarily taken 2D channel is
described by
\begin{align}
\label{eq:boundary}
\omega(x) := a\sin(2\pi x)+ b\, ,
\end{align}
where $b>a$. The sum and difference of the two parameters $a+b$ and
$b-a$ yield  half of the maximal and the minimal width of the
channel, respectively. Moreover, $a$ controls the slope of the
channel walls which in turn determines the one-dimensional diffusion
coefficient $D(x)$.

\begin{figure}[t]
  \centering
  \includegraphics{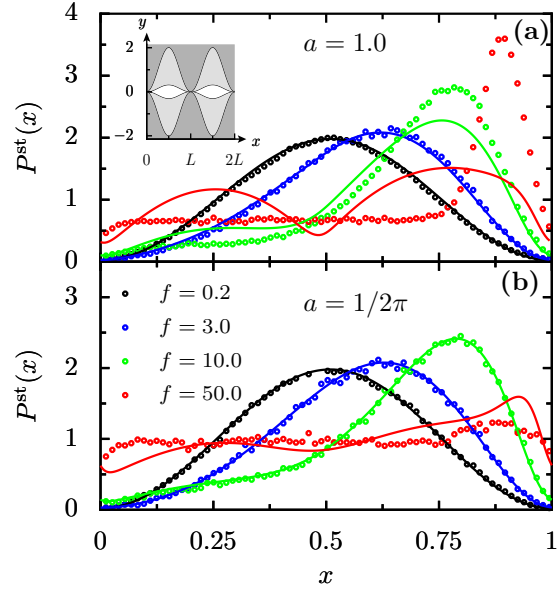}
  \caption{The normalized steady-state probability density of
    particles along the propagation direction is depicted for 
    different force values $f$ (see in panel (b))
    for two different 2D-channels (see inset in panel (a)), 
    with the scaled half-width shape function given by
    $\omega(x)= a \left[\sin(2\pi(x - 0.25)) + 1.02\right]$
    (the shift $-0.25$ ensures that the bottlenecks are located at $0$
    and $1$).  For $a=1$ the maximal and minimal channel widths are
    $4.04$ and $0.04$, respectively; likewise for $a=1/2\pi$ 
    they are $6.43 \cdot 10^{-1}$ and $6.37 \cdot
    10^{-3}$, respectively.
    The Solid lines correspond to the steady-state probability density
    obtained from 1D Fick-Jacob approximation, Eqs.~(\ref{eq:probdist}, 
    \ref{eq:integrals}), and symbols correspond to 2D numerical 
    simulations, see Eq.~(\ref{eq:xprob}).}
\label{fig:probx}
\end{figure}

For the considered channel configuration, cf.
Eq.~(\ref{eq:boundary}), the boundary condition becomes $\omega(x) =
a  \left[\sin(2\pi x) + \kappa \right]$, where $\kappa = b/a = 1.02$
throughout this paper. For $a$ we chose values between 1 and $1/2
\pi$.
In all
cases the width of the widest opening within the channel is larger
by a factor of $100$ than the width at the narrowest opening. One
may therefore do expect rather strong entropic effects for these
channel geometries.

\subsection{Stationary probability densities}

\begin{figure}[t]
\centering
  \includegraphics{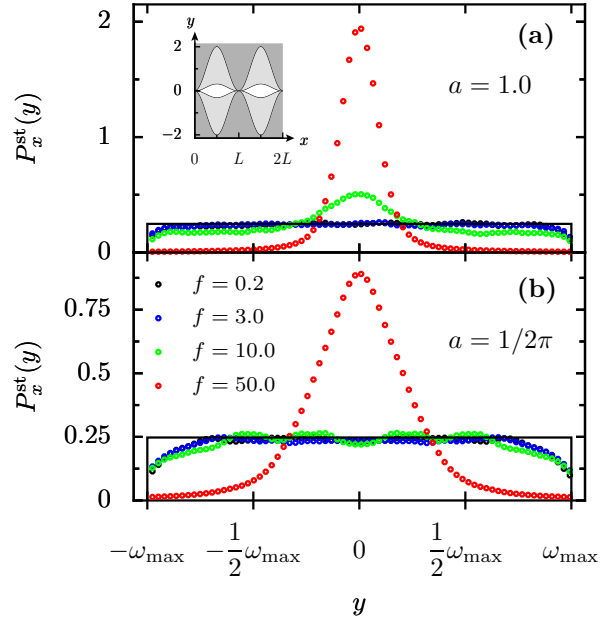}
  \caption{The normalized steady-state probability
    density of particles in $y$-direction $P_{x}^{\mathrm{st}}(y)$,
    cf. Eq.~\eqref{eq:yprob}, taken at the $x$-position of maximal channel
    width, is depicted for different values of the scaling parameter
    $f$ (see in panel (b)) and different channel geometries 
    (see inset in panel (a)) for the boundary function
    $\omega(x)= a \left[\sin(2\pi(x - 0.25)) + 1.02\right]$
    (the shift $-0.25$ ensures that the bottlenecks are located at $0$ and
    $1$). The maximal channel widths for the two structures are:
    $2\omega_{\mathrm{max}}=4.04$ for $a=1$ and
    $2\omega_{\mathrm{max}}=6.43 \cdot 10^{-1}$ for $a=1/2\pi$.
    The symbols correspond to 2D numerical simulations, see Eq.~(\ref{eq:yprob}).
    For large scaled force values $f$ the numerically obtained steady-state
    densities deviate from the uniform, i.e. box-like,
    density (shown by the black solid line). This indicates the failure of
    the equilibration assumption on which the Fick-Jacobs-approximation  relies.}
  \label{fig:disty}
\end{figure}

\noindent We have evaluated the stationary probability density
$P^{\mathrm{st}}(x,y)$, in the long time limit, by mapping all
particle positions onto the primitive cell by translation in
longitudinal channel direction. Consequently, $\int_{0}^{1}\,
\mathrm{d} x \, \int_{-\omega(x)}^{\omega(x)}\, \mathrm{d} y \,
P^{\mathrm{st}}(x,y) = 1$. Fig.~\ref{fig:probx} (solid lines)
depicts the normalized steady-state probability density in
$x$-direction, for various scaling parameter values, derived from
the reduced one-dimensional FJ result, Eq.~(\ref{eq:steadystate}),
and is compared with the  numerical simulations for the exact
expression:
\begin{align}
  \label{eq:xprob}
  P^{\mathrm{st}}(x):=
\frac{\displaystyle \int_{-\omega(x)}^{\omega(x)} P^{\mathrm{st}}(x,y)\,\,\mathrm{d}y}
{\displaystyle\int_{0}^{1}\mathrm{d}x \displaystyle \int_{-\omega(x)}^{\omega(x)} \,
    P^{\mathrm{st}}(x,y)\,\,\mathrm{d}y}\, .
\end{align}
Note, that steady-state marginal density $P^{\mathrm{st}}(x)$ is
normalized on the primitive cell.

At small scaling parameter values $f$ the 1D steady-state
density given by Eq.~(\ref{eq:probdist}) is in very good
agreement with those obtained from numerical simulations, see
Fig.~\ref{fig:probx}. This holds  true for rather arbitrary  channel
geometry (not shown). However, the comparison fails for large
$f$-values of the scaling parameter or for more winding structures
corresponding to larger $a$-values. When increasing the force, the
maximum of $P^{\mathrm{st}}(x)$ is shifted towards the exit of the
cell and the particles mostly accumulate {\itshape in front of} the
bottleneck, see Fig.~\ref{fig:probx}(a), and the 1D kinetic
description starts to fail in that forward bottleneck $x$-region.
However, by decreasing $a$ of the geometric channel shape function
the accuracy of the FJ approximation considerably improves up to
very large force values $f$, see Fig.~\ref{fig:probx}(b).

As a common feature one observe that for the two chosen  geometric
structures that in the large force regime the numerically obtained
$P^{\mathrm{st}}(x)$ is essentially constant over a wide range of
$x$-values, indicating a minor influence of the shape of the
structure on the dynamics of the laterally forward-forced particles.
In this situation, the thermal noise plays a minor role and the
deterministic dynamics (with diffusion set to zero) of the diffusive
equation provides a good starting point. Put differently, at strong
longitudinal driving strength the correction in the diffusion
coefficient leading to a spatial dependency, i.e. $D(x)$,
overestimates the role of the entropic effects and consequently the
FJ approximation starts failing over extended $x$-regimes.

The reasons for the failure of the FJ approximation for large forces
becomes obvious when checking the equilibration assumption in
transverse channel direction. From our simulations, we can actually
analyze the validity of the hypotheses of equilibration in the
transverse direction on which the FJ description relies. A detailed
analysis is provided by testing the normalized steady-state
probability density in the transverse direction at a given
$x$-position, i.e.
\begin{align}
  \label{eq:yprob}
  P_{x}^{\mathrm{st}}(y):=\frac{P^{\mathrm{st}}(x,y)}
{\displaystyle \int_{-\omega(x)}^{\omega(x)} \,
  P^{\mathrm{st}}(x,y) }\, \mathrm{d}y\,  .
\end{align}

In Fig.~\ref{fig:disty}, we depict the steady-state probability
density at the position of maximal channel width.
For small values of the scaling parameter $f$, the
$P_{x}^{\mathrm{st}}(y)$ is very flat, indicating an almost ideal
homogeneous equilibration in the transverse direction, as required
by the FJ approximation scheme. However, at large force
strengths $f$ the Brownian particles concentrate along the axis of
the channel with $y=0$. In this situation, the assumption of
equilibration along the transverse direction fails and the
density peaks around the $y=0$ value. The particles can only
feel the presence of the boundaries when they are close to the
bottlenecks. Hence, in the limit of very large force values, the
influence of the entropic barriers practically disappears.

\subsection{Nonlinear mobility}

The average particle current was derived from an
ensemble-average  using $3\cdot 10^{4}$ trajectories:
\begin{align}
  \label{eq:current-num}
  \langle \dot{x}\rangle &=\lim_{t\to \infty}  \frac{\langle x(t)
    \rangle}{t}\, .
\end{align}

\begin{figure}[t]
  \centering
  \includegraphics{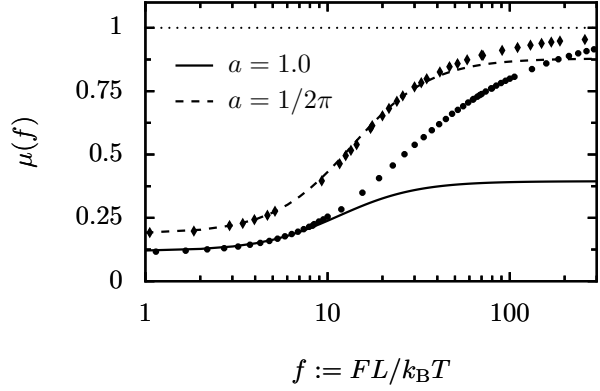}
  \caption{The numerically simulated (symbols) and
    analytically calculated (cf. Eq.~(\ref{eq:nonlinearmobility})
    -- lines)  dependence of the scaled nonlinear
    mobility $\mu(f)$ {\it vs.} the dimensionless force $f=FL/k_{\rm B}T$ is
    depicted for two 2D channel geometries.
    For both channels the scaled half-width is given by
    $\omega(x)= a \left[\sin(2\pi x) + 1.02\right]$;
    $a=1:$ circles and solid line,
    $a=1 / (2\pi):$ diamonds and dashed line.
    The dotted line indicates the deterministic limit $\mu(f) =
    {\langle\dot x \rangle}/f = 1$.}
\label{fig:mobility}
\end{figure}

Fig. \ref{fig:mobility} shows the nonlinear mobility as a function
of the scaling parameter $f$. We note that the transport in
one-dimensional periodic {\it energetic} potentials distinctly
differs  from the one occurring in one-dimensional periodic systems
in presence of {\it entropic} barriers (Reguera {\itshape et al.}, 2006). 
The fundamental difference
lies in the temperature dependence of these barrier shapes.
Decreasing the temperature in an energetic periodic potential
decreases the transition rates from one cell to the neighboring one
by decreasing the Arrhenius factor $\exp\{-\Delta V/(k_{\rm B}T)\}$
where $\Delta V$ denotes the activation energy necessary to proceed
a period (H\"anggi {\it et al.}, 1990).
Hence decreasing  the temperature yields a decreasing nonlinear
mobility. For a one-dimensional periodic system with an entropic
free energy (or entropic potential of mean force), a decrease of
temperature results, however, in an  increase of the dimensionless
force parameter $f$ and consequently in a monotonic  increase of the
nonlinear mobility, cf. Fig. \ref{fig:mobility}.

The dependence of the dynamics on the geometry parameter $a$ nicely
reflects the entropic effects on the mobility: A channel with a
larger $a$-value has wider openings and thus provides more
configuration space where the particle can sojourn. This longer
residence time within a period of the channel diminishes the
throughput and consequently the mobility. This is corroborated by
the results of our calculations depicted in Fig. \ref{fig:mobility}.
For all values of $f$, an increase in value of $a$ leads to a
decrease in the mobility. This holds not only in regimes for which
the FJ equation is valid, but also for large values of $f$ where the
approximation  fails. For very large values of the scaling parameter
$f$ the nonlinear mobility approaches  the value $1$, i.e., it
agrees with the deterministic strong driving  limit.

By means of the nonlinear mobility a detailed comparison between 2D
simulation results and the analytic results, cf.
Eq.~(\ref{eq:nonlinearmobility}), enables one to determine validity
criteria for the  FJ approximation, for further details see in Ref. (Burada {\it et al.},
2007; Burada {\it et al.}, 2008$b$).

\section{Effective diffusion and the Sutherland-Einstein relation}

\begin{figure}[t]
  \centering
  \includegraphics{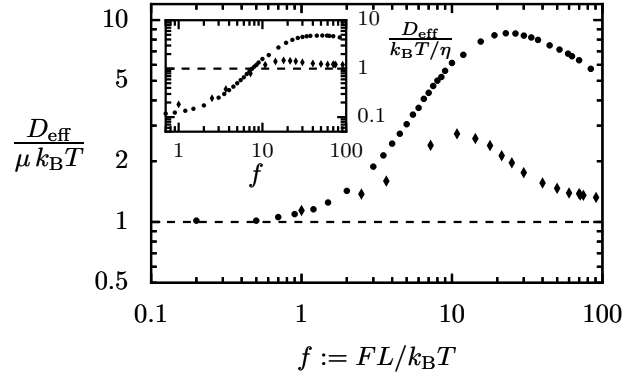}
  \caption{The ratio of the effective Diffusion $D_{\mathrm{eff}}$ and
    nonlinear mobility $\mu$ times
    the thermal energy $k_{\mathrm{B}} T$ is depicted as a function
    of the scaling parameter $f$ for two channel
    geometries: $\omega(x) = a [\sin(2\pi x) + 1.02]$ with $a=1$ (circles)
    and $a=1/2\pi$ (diamonds). The dashed horizontal line at $1$ indicates the
    validity of  the Sutherland-Einstein relation in this non-equilibrium situation: 
    $D_{\mathrm{eff}} / (\mu \, k_{\mathrm{B}}T) = 1$. 
    A deviation from this line consequently marks the breakdown
    of this relationship. The inset depicts the ratio
    of the effective diffusion $D_{\mathrm{eff}}$ and bulk diffusion
    constant, being $k_{\mathrm{B}}T / \eta$. A ratio larger
    than $1$ (dashed horizontal line) indicates a characteristic 
    enhancement of the effective $x$-diffusion.}
  \label{fig:einsteinrel}
\end{figure}

A validity of a nonlinear Sutherland-Einstein relation implies that in physical units
we can relate the nonlinear mobility $\mu(F)$ directly to the
nonlinear, effective $x$-diffusion $D_{eff}(F)$, reading

\begin{equation}
  \label{eq:einsteinrel}
  D_{\mathrm{eff}}(F) = \mu(F) \,  k_{\mathrm{B}} T \, ,
\end{equation}
Put differently, the effective diffusion coefficient
$D_{\mathrm{eff}}$ for the
diffusive spreading along the longitudinal channel direction would
then solely be determined by the nonlinear
mobility  discussed above and the environmental temperature $T$.

A validity of this relation would then imply a monotonic increase
towards the entropic-free diffusion limit, i.e.  $ D_{eff} =k_\mathrm{B}T/ \eta$. The latter
is being  approached in the strong forcing limit
where entropic effects cease to play a significant role. Such a
monotonic behavior, however, is not observed from the
numerical simulations for the effective $x$-diffusion coefficient 
(Burada {\it et al.}, 2008$b$; Reguera {\it et al.}, 2006). 
It is defined as the ratio  between the  asymptotic
behavior of the variance of the position variable and the elapsed time $t$; i.e.,
\begin{equation}
   \label{eq:diffusion-num}
D_{\text{eff}} = \lim_{t \to \infty} \frac{\langle x^{2}(t) \rangle -\langle
  x(t) \rangle^{2}}{2t} \, .
\end{equation}
Interestingly, the dependence of the effective
diffusion coefficient on the scaling parameter exhibits a bell-shaped
behavior, cf. inset of Fig.~\ref{fig:einsteinrel}, thus indicating a
failure of the Sutherland-Einstein relation in this moderate-to-strong
driving regime. This break down of the Sutherland-Einstein
relation can also be detected within the FJ description (not shown in
Fig.~\ref{fig:einsteinrel}): The FJ approximation for 
this effective $x$-diffusion as well yields a
non-monotonic dependence of the effective $x$-diffusion coefficient
on the scaling parameter $f$ exhibiting a peak value exceeding the
bulk diffusion coefficient $D_{0}= \eta/k_B T$ 
(Burada {\it et al.}, 2008$b$; Reguera {\it et al.}, 2006). 

For a detailed
comparison, we depict the ratio of numerically obtained
$D_{\mathrm{eff}}$ and ($\mu \, k_{\mathrm{B}} T$) in
Fig.~\ref{fig:einsteinrel}. Surprisingly, it turns out that such a
Sutherland-Einstein relation, Eq.~(\ref{eq:einsteinrel}), holds true
in terms of the effective mobility in the small forcing limit  $f \to
0$ ; i.e. in the linear response regime. It increasingly fails,
however, for increasing bias strength $F$. At very strong bias, i.e.,
$f \to \infty$ the biased diffusion becomes effectively ''free'' from
entropic effects and expectedly approaches the  free  limit, given by
$k_\mathrm{B} T /\eta$, which renders
the original, linear Sutherland-Einstein result in terms of the
$F$-independent mobility $\mu= 1/\eta$. Put differently, the influence
of entropic barriers caused by the bottlenecks becomes negligible at
strong bias. Vice versa, the bell-shaped behavior of the ratio
depicted with Fig.~\ref{fig:einsteinrel} reflects the fact that this
effective diffusion is not increasing monotonically but rather
exhibits an enhancement of effective diffusion at moderate bias (or
scaling) values $f$, c.f. in the inset of Fig.~\ref{fig:einsteinrel}.

\section{Conclusions}

In summary, we demonstrated the applicability of the
equilibration approximation in describing biased diffusive transport
occurring in narrow, irregularly shaped one-dimensional channel
structures. The Fick-Jacobs (FJ) description which relies on the
equilibration assumption allows for a treatment of the dynamics
within an effective one-dimensional kinetic equation of the
Smoluchowski form. Bottlenecks and other confining restrictions of
available configuration space yield within this approximation an
effective  1D diffusion equation exhibiting  entropic barriers. Due
to the intrinsic  temperature dependence of the underlying entropic
free energy contribution one finds for the transport phenomena in
periodic channels possessing varying cross sections features that
are radically different from conventional transport occurring in
energetic periodic potential landscapes.

The most striking
difference between these two physical situations is  that for a
fixed channel geometry the dynamics is  characterized by a single
scaling parameter $f= FL/(k_{\rm B}T)$ which combines the external
force $F$ causing a drift, the period length $L$ of the channel, and
the thermal energy $k_{\rm B}T$.  The latter presents  a measure of
the strength of the acting fluctuating thermal forces. This leads to
an opposite temperature dependence of the mobility: While the
mobility of a particle in an energetic potential increases with
increasing temperature the mobility of a particle undergoing biased
diffusion in an irregular channel  decreases. The incorporation of
the spatial variation of the channel width in terms of an entropic
free energy contribution  allows for a quantitative understanding of
the dependence of the transport properties, like the nonlinear
mobility, on parameters like force strength, channel topology or
temperature. Moreover, the lateral steady-state probability  densities
$P^{\mathrm{st}}(x)$ can be
evaluated in analytical closed form within the reduced kinetic
FJ approximation, see in the  Appendix.

Such an effective one-dimensional reduction of a
complex diffusion dynamics with intricate boundary conditions at the
confining walls certainly proves useful and  beneficial for the
quantitative description, design and control of diffusive transport 
along tortuous pores
and alike. The latter situation  dictates the stochastic
far-from-equilibrium transport in a  great variety of biological and
structured synthetic pores and confining cavities, such as
buckyballs, zeolites and alike. As an example, this FJ approximation
has successfully been used in describing the phenomenon of
Stochastic Resonance (H\"anggi, 2002; Gammaitoni {\it et al.}, 1998) 
in a 2D system exhibiting an entropic barrier (Burada {\it et al.}, 2008$a$).

\bigskip
{\small This work has been supported by the Volkswagen Foundation
(project I/80424, P.H.), the DFG via research center, SFB-486,
project A10 (G.S., P.H.), and via the DFG project no. 1517/26-1
(P.S.B., P.H.), and by the German Excellence Initiative via the
\textit {Nanosystems Initiative Munich} (NIM) (P.H., P.S.B.).}

\begin{appendix}{Steady-state Current and Probability Density}
\label{app:steadystate}

In this appendix we derive the steady-state solution for
the effective, one-dimensional dimensionless FJ
(Smoluchowski-type) equation, Eq.~(\ref{eq:fickjacobs}),
\begin{align}
  \label{eq:1Dkin}
  \frac{\partial}{\partial t} P(x,t) =
  \frac{\partial}{\partial x} \,D(x) \,e^{-A(x)}
  \frac{\partial}{\partial x} e^{A(x)} P(x,t) \, ,
\end{align}
where $A(x)$ denotes the free energy function, $A(x) = -f\, x -  \ln
\omega(x)$, with $\omega(x+1) = \omega(x)$. Eq.~(\ref{eq:1Dkin})
results from the probability continuity equation:
\begin{align}
  \label{eq:continuity}
  \frac{\partial}{\partial t}P(x,t) = - \frac{\partial}{\partial x}J(x,t) \, ,
\end{align}
with the probability current $J(x,t)$ reading:
\begin{align}
  \label{eq:j1}
  J(x,t)=\,-\,D(x) \,e^{-A(x)}
  \frac{\partial}{\partial x} e^{A(x)} P(x,t) \, .
\end{align}

In the case of a tilted periodic potential, i.e. $A(x+1)=A(x)-f$, it
is convenient to define the reduced probability density and the
corresponding reduced current; i.e.,
\begin{align}
  \label{eq:redprobdens}
  \hat{P}(x, t) &= \sum_n P(n+x, t) \, ,\\
  \hat{J}(x, t) &= \sum_n J(n+x, t) \, ,~n \in \mathbb{Z}  \, .
\end{align}

By definition these functions are periodic with  periodicity $L=1$,
$\hat{P}(x+1, t) = \hat{P}(x, t)$, $\hat{J}(x+1, t) = \hat{J}(x,t)$.
The same holds true for the spatial dependent diffusion coefficient,
i.e. $D(x+1) = D(x)$. Moreover, $\hat{P}(x, t)$ and $\hat{J}(x, t)$
obey the continuity equation, Eq.~(\ref{eq:continuity}), and
$\hat{P}(x,t)$ is normalized on an interval $(x, x + 1)$, provided
that $P(x, t)$ is normalized, i.e. $\int_{-\infty}^{+\infty} P(x,
t)\,\mathrm{d}x = 1$.

In the {\it steady-state} limit the probability current assumes a
constant, i.e., $\hat{J}(x,t) \rightarrow \hat{J}$. Thus
Eq.~(\ref{eq:j1}) becomes
\begin{align}
  \label{eq:j2}
  \hat{J}= \,-\,D(x) \,e^{-A(x)}
  \frac{\partial}{\partial x} e^{A(x)} \hat{P}^{\rm st}(x) \, .
\end{align}
Multiplying both sides of Eq.~(\ref{eq:j2}) by $e^{A(x)}/D(x)$, and
integrating over a period $L=1$  we obtain
\begin{align}
  \label{eq:j3}
  \hat{J}\,\int_{x}^{x+1}\, \frac{e^{A(x')}}{D(x')}\,
  \mathrm{d}x' =
  -\,\int_{x}^{x+1} \frac{\partial}{\partial x'} e^{A(x')}\,
  \hat{P}^{\rm st}(x')\, \mathrm{d} x' \, ,
\end{align}
which simplifies to
\begin{align}
\label{eq:j5}
  \hat{J}\,\int_{x}^{x+1}\, \frac{e^{A(x')}}{D(x')}\,
  \mathrm{d}x' =
 \hat{P}^{\mathrm{st}}(x) \left( 1\, - \, e^{-f} \right) e^{A(x)}\, .
\end{align}
Upon rearranging the terms on the right hand side and integrating
once more over a period, i.e. from $0$ to $1$, we find the first result
\begin{align}
  \label{eq:j7}
  \hat{J} = \frac{\left(1-e^{-f}\right)}
  {\displaystyle \int_{0}^{1}\,e^{-A(x)}\,\mathrm{d}x
    \displaystyle
    \int_{x}^{x+1}\,\frac{e^{A(x')}}{D(x')}\,\mathrm{d}x'} \, .
\end{align}
Hereby,  we made use of the normalization condition of the stationary
probability, i.e., $\int_{0}^{1}
\hat{P}^{\mathrm{st}}(x)\,\mathrm{d}x = 1$. The general relation
between the steady-state probability current and the steady-state
average particle current ($\langle \dot{x} \rangle$) is
\begin{align}
\label{eq:current}
\langle \dot{x} \rangle = \int_{0}^{1} \hat{J}\, \mathrm{d}x
\end{align}
which implies that $\langle \dot{x} \rangle = \hat{J}$. Thus, the
transport  current is given by the first main result, reading:
\begin{align}
  \label{eq:j8}
  \langle \dot{x} \rangle = \frac{\left(1-e^{-f}\right)}
  {\displaystyle \int_{0}^{1}\,e^{-A(x)}\,\mathrm{d}x
    \displaystyle \int_{x}^{x+1}\,\frac{e^{A(x')}}{D(x')} \,\mathrm{d}x'} \, .
\end{align}

By substituting Eq.~(\ref{eq:j7}) back into the Eq.~(\ref{eq:j5}) we
 obtain for the steady-state probability density  in
$x$-direction the second main result:
\begin{align}
  \label{eq:steadystate}
  \hat{P}^{\mathrm{st}}(x) =
  \frac{e^{-A(x)}\, \displaystyle \int_{x}^{x+1}\,
\frac{e^{A(x')}}{D(x')}\,\mathrm{d}x'}
{\displaystyle \int_{0}^{1}\,e^{-A(x)}\,\mathrm{d}x
  \displaystyle \int_{x}^{x+1}\,\frac{e^{A(x')}}{D(x')}\,\mathrm{d}x'}\;\;.
\end{align}

\end{appendix}


\section*{References}

\begin{thedemobiblio}{20}

\item Astumian, R.D. \& H\"anggi, P. 2002 Brownian Motors.
{\it Physics Today} {\bf 55}, 33-39.

\item Barrer, R.M. 1978 
  {\it Zeolites and Clay Minerals as Sorbents and Molecular Sieves},
  London: Academic Press

\item Beckurts, K.H. \& Wirtz, K. 1964 
  {\it Neutron Physics}, Berlin: Springer

\item Berezhkovskii, A.M. \& Bezrukov, S.M. 2005
  Optimizing Transport of Metabolites through Large Channels:
  Molecular Sieves with and without Binding.
  {\it Biophys. J.} {\bf 88}, L17-L19.

\item Burada, P. S., H\"anggi, P., Marchesoni, F., Schmid, G. \&
  Talkner, P. 2009 Diffusion in confined geometries.
  {\it ChemPhysChem.} {\bf 10} 45-54.

\item Burada, P.S., Schmid, G., Reguera, D., Rub\'i, J.M.
  \& H\"anggi, P. 2007
  Biased diffusion in confined media:
  Test of the Fick-Jacobs approximation and validity criteria.
  {\it Phys. Rev.} {\bf E75}, 051\,111-1-051\,111-8.

\item Burada, P.S., Schmid, G., Reguera, D., Vainstein, M.H., Rub\'i, J.M.
  \& H\"anggi, P. 2008a
  Entropic stochastic resonance.
  {\it Phys. Rev. Lett.} {\bf 101} 130\,602-1-130\,602-4.

\item Burada, P.S., Schmid, G., Talkner, P., H\"anggi, P.,
  Reguera, D. \& Rub\'i, J.M. 2008b
  Entropic particle transport in periodic channels.
  {\it Biosystems} {\bf 93} 16-22.

\item Derenyi, I. \& Astumian, R.D. 1998
  ac separation of particles by biased Brownian motion in a two-dimensional sieve.
  {\it Phys. Rev.} {\bf E58}, 7\,781-7\,784.

\item Gammaitoni, L.,  H\"anggi, P., Jung, P. \& Marchesoni, F. 1998 
  Stochastic resonance.
  {\it Rev. Mod. Phys.} {\bf 70}, 223-288.

\item H\"anggi, P. 2002
  Stochastic Resonance in Biology.
  {\it ChemPhysChem.} {\bf 3}, 285-290.

\item H\"anggi, P., \& Marchesoni, F. 2005
  Introduction: 100 Years  of Brownian motion.
  {\it Chaos} {\bf 15}, 026\,101-1-026\,101-5;  
  see in particular reference {\it 3} therein.

\item H\"anggi, P., \& Marchesoni, F. 2009
  Artificial Brownian motors: Controlling transport on the nanoscale.
  {\it Rev. Mod. Phys.} {\bf 81}, 1-55.

\item H\"anggi, P., Talkner, P. \& Borkovec, M. 1990
  Reaction Rate Theory: Fifty Years After Kramers.
  {\it Rev. Mod. Phys.} {\bf 62}, 251-342.

\item H\"anggi, P. \& Thomas, H. 1982
  Stochastic Processes: Time-Evolution, Symmetries and Linear Response.
  {\it Phys. Rep.} {\bf 88}, 207-319.

\item Hille, B. 2001 
{\it Ion Channels of Excitable Membranes}, Sunderland: Sinauer.

\item Jacobs, M.H. 1967
{\it Diffusion Processes}, New York: Springer.

\item Kalinay, P. \& Percus, J.K. 2006
  Corrections to the Fick-Jacobs equation.
  {\it Phys. Rev.} {\bf E74}, 041\,203-1-041\,203-6.

\item Kettner, C., Reimann, P., H\"anggi, P. \& M\"uller, F. 2000
  Drift ratchet.
  {\it Phys. Rev.} {\bf E61}, 312-323.

\item Kosinska, I.D., Goychuk, I., Kostur, M., Schmid, G. \& H\"anggi, P. 2008
  Rectification in synthetic conical nanopores:
  a one-dimensional Poisson-Nernst-Planck modeling.
  {\it Phys. Rev.} {\bf E77}, 031\,131-1-031\,131-10.

\item Liu, L., Li, P. \& Asher, S.A. 1999
  Entropic trapping of macromolecules by mesoscopic
  periodic voids in a polymer hydrogel.
  {\it Nature} {\bf 397}, 141-144.

\item Matthias, S. \& M\"uller F. 2003
  Asymmetric pores in a silicon membrane
  acting as massively parallel brownian ratchets.
  {\it Nature} {\bf 424}, 53-57.

\item Mazo, R.M. 2002
  {\it Brownian Motion: Fluctuations, Dynamics and Applications},
  Oxford: Clarendon Press.

\item M\"uller, F., Birner, A., Schilling J., G\"osele, U.,
  Kettner, C., \& H\"anggi, P. 2000
  Membranes for Micropumps from Macroporous Silicon.
  {\it phys. stat. sol. (a)} {\bf 182}, 585-590.

\item Nixon, G.I. \& Slater, G.W. 2002
  Saturation and entropic trapping of monodisperse polymers in porous media.
  {\it J. Chem. Phys.} {\bf 117}, 4\,042-4\,046.

\item Purcell, E.M. 1977
  Life at low Reynolds number.
  {\it Am. J. Phys.} {\bf 45}, 3-11.

\item Reguera, D. \& Rub\'i, J.M. 2001
  Kinetic equations for diffusion in the presence of entropic barriers.
  {\it Phys. Rev.} {\bf E64}, 061\,106-1-061\,106-8.

\item Reguera, D., Schmid, G., Burada, P.S.,
  Rub\'i, J.M., Reimann, P. \& H\"anggi, P. 2006
  Entropic transport: Kinetics, scaling and control mechanisms.
  {\it Phys. Rev. Lett.} {\bf 96}, 130\,603-1-130\,603-4.

\item Reimann, P. \& H\"anggi, P. 2002 
  Introduction to the Physics of Brownian Motors.
  {\it Appl. Physics} {\bf A75}, 169-178.

\item Risken, H. 1989
  {\it The Fokker-Planck equation}, 2nd edn. Berlin: Springer.

\item Siwy, Z., Kosinska, I.D., Fulinski, A. \& Martin, C.R. 2005
  Asymmetric Diffusion through Synthetic Nanopores.
  {\it Phys. Rev. Lett.} {\bf 94}, 048\,102-1-048\,102-4.

\item Volkmuth, W.D. \& Austin, R.H. 1992
  DNA electrophoresis in microlithographic arrays.
  {\it Nature} {\bf 358}, 600-602.

\item Zwanzig, R. 1992
  Diffusion Past an Entropic Barrier.
  {\it J. Phys. Chem.} {\bf 96}, 3\,926-3\,930.

\end{thedemobiblio}


\end{document}